\documentclass[12pt]{article}

\usepackage{newtxtext,newtxmath}

\usepackage{graphicx}
\usepackage[normalem]{ulem}

\usepackage[letterpaper,margin=1in]{geometry}

\renewenvironment{abstract}
	{\quotation}
	{\endquotation}

\date{}

\makeatletter
\renewcommand{\fnum@figure}{\textbf{Figure \thefigure}}
\renewcommand{\fnum@table}{\textbf{Table \thetable}}
\makeatother

\usepackage{scicite}

\usepackage{url}

\def\scititle{
	Dynamics of disordered quantum systems with two- and three-dimensional tensor networks
}
\title{\bfseries \boldmath \scititle}

\author{
	Joseph~Tindall$^{1\ast}$,
	Antonio~Francesco~Mello$^{1, 2}$,
	Matthew~Fishman$^{1}$, \and
    E.~Miles~Stoudenmire$^{1}$,
    Dries~Sels$^{1,3\ast}$\and
	\small$^{1}$Center for Computational Quantum Physics, Flatiron Institute, New York, New York 10010, USA.\and
	\small$^{2}$International School for Advanced Studies (SISSA), via Bonomea 265, 34136 Trieste, Italy.\and
    \small$^{3}$Department of physics, Boston University, 590 Commonwealth Avenue, Boston, Massachusetts 02215, USA.\and
	\small$^\ast$Corresponding author. Email: jtindall@flatironinstitute.org, dsels@bu.edu\and
}

\begin{document} 

\maketitle

\begin{abstract} \bfseries \boldmath
Large scale quantum annealing dynamics of Ising spin glasses were recently implemented on D-Wave's Advantage$2$ system on a range of lattices. Following extensive comparison to existing numerical methods, these experiments were claimed to be beyond the reach of classical computation. Here, we simulate these spin glass models with lattice-specific tensor networks, using belief propagation (BP) to keep up with the entanglement generated during the time evolution and then extracting expectation values with more sophisticated variants of BP. We find that state-of-the-art accuracies can be achieved with modest computational resources. Moreover, our results are scalable in both two and three dimensions, which we leverage to verify universal Kibble-Zurek physics on systems involving hundreds of qubits.
\end{abstract}

\noindent
Dynamical simulations are fundamental for understanding the non-equilibrium correlated states of matter which emerge on transient timescales and are routinely being realized in quantum devices thanks to steady progress in quantum technologies~\cite{Xiao2021, Mohsin2024, Keesling2019, Buzzi2020, Frey2022, Qiujian2021, Smith2016}.
    
    Although a wide variety of classical computational methods have been proposed for simulating non-equilibrium quantum systems \cite{Schiro2010, Kota2015, Mingru2020, Kaelan2023, Gutierrez2022, Kloss2020, Paeckel2019, Wurtz18, Reyhaneh2020}, these methods often suffer from severe bottlenecks which make them difficult to scale up. Approaches involving variational ans\"atze such as neural quantum states (NQS), for instance, suffer from numerical stability issues when inverting the quantum geometric tensor, limiting the time horizon \cite{Sinibaldi2023}. The intrinsic one-dimensional structure of matrix product states (MPS), meanwhile, typically makes them a poor ansatz for large, correlated, two- and three-dimensional systems, whereas higher-dimensional tensor networks  often require costly computational methods for optimization and contraction, limiting the amount of entanglement that can be captured.
    
     These difficulties suggest that controllable quantum devices, such as digital quantum processors, quantum annealers or analogue quantum simulators may be the best tools for accurately performing scalable quantum simulations, despite their inherent noise. A number of notable experiments have put forward substantial evidence to this effect~\cite{Arute2019,Morvan2024, king2024, Sepehr2021, Kim2023}. 

     Prominent examples of experiments on programmable, digital processors include random unitary circuits \cite{Arute2019, Morvan2024}. These provide strong evidence of quantum advantage, although their practical significance and the exact quantum speedups remain unsettled \cite{Zhang2022, Gray2021hyperoptimized, Ayral2023,Schuster2025}. More physically motivated circuits \cite{Kim2023, Robledo2025} have also been considered; by contrast, these are often susceptible to classical simulations which can leverage the physical structure in the system to efficiently obtain high-accuracy results \cite{Tindall2024, Begusic2024, Rudolph2023, rudolph2025simulatingsamplingquantumcircuits}.
    
     Recently, the quantum annealing dynamics of a disordered spin model were faithfully realized on D-Wave's Advantage$2$ quantum annealer on a range of two, three, and infinite-dimensional lattice structures~\cite{king2024}. An exhaustive comparison to a range of variational classical simulation methods was performed---including the aforementioned ans\"atze like neural quantum states, matrix product states, and projected entangled pair states (PEPS). The computational difficulties present in these methods were apparent and led to the conclusion that classical methods required exponential resources in the system size. The quantum device thus held computational ``quantum advantage" over them in a physical setup---a long sought-after goal in quantum simulation.

     \begin{figure*}[t!]
    \centering
    \includegraphics[width = \textwidth]{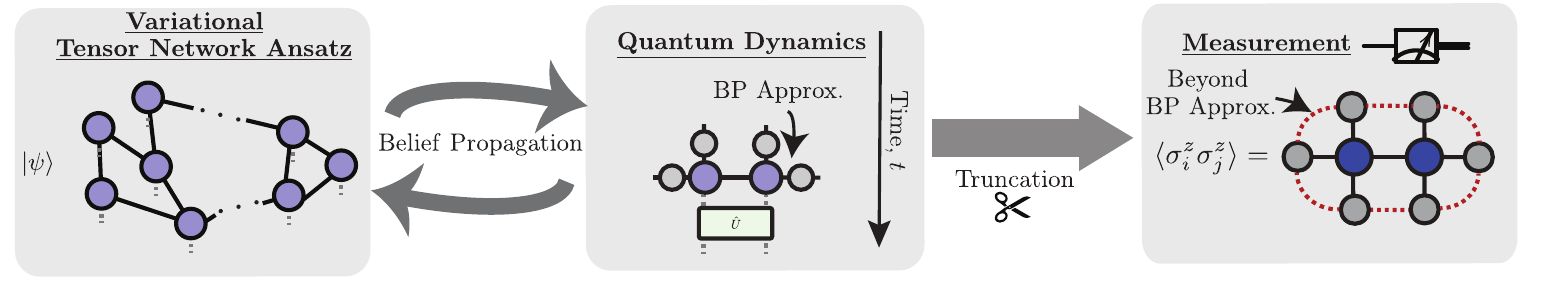}
    \caption{\textbf{Classical approach to quantum simulation.} To simulate the dynamics induced by the quantum spin glass Hamiltonian in Eq.~\ref{Eq:Hamiltonian} on cylindrical, dimerized cubic, and diamond cubic lattices, the wavefunction is encoded as a tensor network (an interconnected network of tensors) whose structure matches the system's underlying geometry. This quantum state is time-evolved via a belief propagation-based simple update scheme, allowing for efficient compression of the shared indices between the tensors in the network to a desired maximum size. 
    Measurements of local and non-local observables are taken intermittently at designated measurement times via corrections of belief propagation, which can account for correlations among environment degrees of freedom. If necessary, a  (belief propagation-based) truncation of the tensors is performed prior to this measurement to enable the efficient use of these more controlled, but computationally more expensive, contraction schemes.}
    \label{fig:MainFigure}%
\end{figure*}
    
     In this work we adopt a belief propagation (BP)-based approach (see Fig.~\ref{fig:MainFigure}) to show that the quantum annealing dynamics of disordered Ising models can be accurately and efficiently captured with a tensor network ansatz whose structure matches the underlying lattice.  We simulate the same glassy two- and three-dimensional systems as in Ref.~\cite{king2024} and show that state-of-the-art accuracies can be achieved with resources that, for a given annealing time, scale only linearly in the system size and are readily available with current classical computing hardware. In the case of cylindrical and diamond lattices our simulations reach accuracies which are well beyond those of the quantum annealer and appear to be controlled with increasing system size; in the case of the dimerized cubic lattice the accuracies we reach are comparable to the annealer. 

     On both cylindrical and diamond lattices we simulate the universal physics associated with dynamically crossing a phase transition on hundreds of qubits. This allows us to extract the associated critical exponents in agreement with literature, demonstrating that tensor networks are far more capable of simulating quantum dynamics in two and three dimensions than previously expected. 
     
\subsection*{Model and Methodology}
We consider the quantum dynamics induced by the following time-dependent Hamiltonian
\begin{equation}
    H(s) = \mathcal{J}(s)\sum_{\langle i, j \rangle}J_{ij}\sigma^{z}_{i}\sigma^{z}_{j} + \Gamma (s)\sum_{i \in \Lambda}\sigma^{x}_{i},
    \label{Eq:Hamiltonian}
\end{equation}
where $s = \frac{t}{t_{a}}$ is a renormalized time parameter, $t_{a}$ is the total annealing time, $\mathcal{J}(s)$ and $\Gamma(s)$ are functions set by an annealing schedule, and $\{J_{ij}\}$ are a set of couplings between the nearest neighbors of the underlying lattice $\Lambda$. We consider ``high-precision'' couplings drawn independently and uniformly from the set of all binary fractions $a / 128$ with $a \in \mathbb{Z}$ and $-128 \leq a \leq 128$. We use the same annealing schedule as was used for benchmarking in Ref. \cite{king2024}.

We consider a quench starting from the ground state of $H(0)$ from $s = 0$ ($t = 0$) to time $s = 1$ ($t = t_{a}$), with the total annealing time $t_{a}$ controlling the adiabaticity of the evolution. At time $s = 1$ we calculate two-point correlators $\langle \sigma^{z}_{i}\sigma^{z}_{j} \rangle$ for both local and non-local pairs of qubits in the lattice, building up a picture of the spin glass order induced by the couplings $J_{ij}$.

To model the quench we adopt a classical simulation approach (Fig.~\ref{fig:MainFigure}). We use a tensor network ansatz for the many-body wavefunction whose structure reflects the underlying lattice of interest. A tensor network can be understood as a compressed format for the many-body wavefunction, consisting of a network of tensors — one for each qubit — connected by indices which mediate the entanglement between the qubits. Increasing the size of these indices (known as the ``bond dimension'') increases the expressivity of the ansatz.

In order to apply gates to a tensor network and extract information from it (e.g.~measure observables) it is in general necessary to take the derivative of the tensor network representing the norm of the wave function $\langle \psi \vert \psi \rangle$ with respect to some subset of the tensors in $\vert \psi \rangle$ and their conjugates.
 When the original network $\vert \psi \rangle$ contains loops the contractions necessary to compute derivatives are not efficiently computable in general \cite{Schuch2007, HaferKamp2020} and an approximate, `low-rank' form for the derivatives must be found instead.

When applying nearest-neighbor gates (following a second-order Trotterization of the propagator) we use standard BP-based message passing to find an approximate separable (i.e.~rank-$1$) form for this derivative and truncate the virtual bonds conditioned on this approximation \cite{Alkabetz2021, tindallbeliefpropagation, Vidal2004, Jiang2008}. This BP-based ``simple update" method is extremely efficient compared to more controlled update methods \cite{Lubasch2014, Phien2015}. Although it can lead to uncontrolled truncations of the virtual bonds, it still has the property that when no truncation is performed there is no error in the application of the gate \cite{Supplementary}. We find that the efficiency of the BP method is fundamental for capturing the entanglement generated by the annealing circuits studied in this work and thus for maintaining a low overall truncation error given the computational resources available to us. 

Once the entire evolution has been simulated, we apply more controlled contraction methods to measure observables. We find such methods are necessary to handle the small loop sizes in the lattices studied and that standard BP contraction---which has recently been shown to be successful for annealing on tree-like systems \cite{luchnikov2024}---is not sufficiently accurate. If the bond dimension of the network following evolution is too high to  apply the controlled methods, we perform a truncation of the virtual bonds of the network down to a more affordable one via the final BP messages \cite{tindallbeliefpropagation}.

The first of the controlled contraction methods we use is our adaptation of the well-established boundary matrix product state algorithm \cite{Verstraete2004, Lubasch2014} which, conventionally, optimizes a correlated matrix product state (MPS) to represent sequential contractions of rows or columns of a square-lattice tensor network. Here, we use a one-site fitting routine for optimal efficiency and have adapted the method to work on any tensor network which, upon appropriate grouping of the tensors, forms an open boundary or half-periodic planar lattice. 
In this work we focus on a cylindrical tensor network (see Fig.~\ref{fig:DWaveErrorvsBondDimension}) with the matrix product states passed as ``messages'' iteratively around the cylinder until convergence. This is very much in the spirit of the classic BP message passing algorithm but with MPS messages and thus we refer to it as MPS message passing. In Fig.~\ref{fig:DWaveErrorvsBondDimension}B and the Supplementary Text \cite{Supplementary}  we show how, for a fixed TNS bond dimension and MPS rank, we can use the MPS message passing method to compute all $\frac{N(N-1)}{2}$ two point correlators $\langle \sigma_{i}^{z} \sigma^{z}_{j} \rangle$ in $\mathcal{O}(N^{2})$ time, where $N$ is the number of spins. 

The second controlled contraction method we use can be applied to any lattice structure and builds off of the recently introduced loop-corrected BP algorithm \cite{evenbly2024} to approximate the contraction of the tensor network $\langle \psi \vert \sigma_{i}^{z} \sigma^{z}_{j} \vert \psi \rangle$. Loop-corrected BP involves viewing the contraction of the tensor network as a sum over ``configurations'', where a configuration is generated by placing either the projector $\mathcal{P}_{e}$ into the BP subspace (formed from the outer product of the relevant message tensors) or its antiprojector $\mathbb{I} - \mathcal{P}_{e}$ on each of the edges $e$ of the network. The sum can then be approximated by enumerating \cite{Johnson1975, Fairbanks2021} and adding up configurations which contain a maximum number $l_{\rm max}$ of antiprojectors \cite{Supplementary}.

\subsection*{Cylindrical Lattice}

\begin{figure*}[t!]
    \centering
    \includegraphics[width =\textwidth]{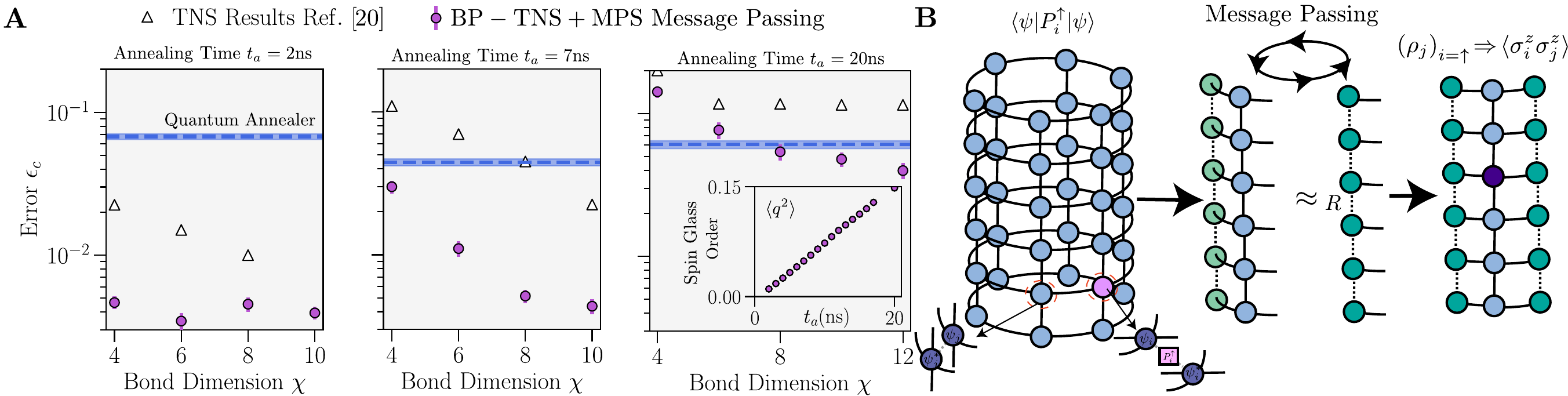}
    \caption{
    \textbf{Cylindrical lattice.} (\textbf{A}) Error $\epsilon_{c}$---see Eq.~\ref{Eq:ErrorMetric}---from two-dimensional tensor network (TN) simulations of a glassy quantum annealing protocol on an $8 \times 8$ cylinder. The same $n = 20$ disorder realizations are used as in Ref.~\cite{king2024}. 
     Error bars correspond to double the standard error on the mean. Orchid data points show results from our BP-based evolution protocol with maximum bond dimension $\chi_{\rm BP} = 32$. We truncate the final state, with BP, to a bond dimension $\chi$ and use cylindrical MPS message passing with MPS dimension $R = 2\chi$ to calculate $\langle \sigma^{z}_{i}\sigma^{z}_{j} \rangle \ \forall i, j$.  Blue dotted line shows the average error from the D-Wave Advantage2 annealer. The shaded region indicates $\pm 2\sigma / \sqrt{n}$. White markers are the average error from the 2D TN simulation in Ref.~\cite{king2024}. The inset shows the value we obtain for the spin glass order parameter $\langle q^{2} \rangle$ versus annealing time. (\textbf{B}) Cylindrical MPS message passing method we use to compute two-point correlators $\langle \sigma^{z}_{i} \sigma^{z}_{j} \rangle$ associated with the $\mathbb{Z}_{2}$ invariant TN $\vert \psi \rangle$. The spin-up projector $P^{\uparrow}_{i}$ is inserted into $\langle \psi \vert \psi \rangle$ and boundary MPS is run around the cylinder until convergence. All two point correlators $\langle \sigma^{z}_{i} \sigma^{z}_{j} \rangle$ with $i$ fixed can then be computed.}
    \label{fig:DWaveErrorvsBondDimension}
\end{figure*}

 We first compare our simulation results to ground truth values obtained using a one-dimensional MPS ansatz and the time-dependent variational principle (TDVP) algorithm \cite{Haegeman2011}. For sufficiently small systems, MPS methods can be used to controllably and accurately compute the time evolution and any desired observables and thus serves as an ideal benchmarking method. 
 
 We utilize the same error metric as in Ref.~\cite{king2024}
\begin{equation}
    \epsilon_{c} = \sqrt{\frac{\sum_{i > j}\left(c_{ij} - \tilde{c}_{ij}\right)^{2}}{\sum_{i > j}\tilde{c}^{2}_{ij}}},
    \label{Eq:ErrorMetric}
\end{equation}
where $c_{ij} = \langle \sigma^{z}_{i} \sigma^{z}_{j} \rangle$ is computed from our simulations and $\tilde{c}_{ij}$ denotes the corresponding ground truth values from a converged MPS simulation. The bond dimension required for the MPS approach to achieve fixed accuracy, however, scales exponentially with system size in two- and three-dimensional setups---meaning it cannot be used at scale \cite{king2024}. 

In Fig.~\ref{fig:DWaveErrorvsBondDimension} we show our results from a cylindrical tensor network of size $8 \times 8$ for several different annealing times. There, we use a maximum bond dimension of $\chi_{\rm BP} = 32$ for the simple BP-based time evolution protocol and then truncate the tensor network down to the final bond dimension $\chi$ with BP and perform MPS message passing with MPS rank $R = 2 \chi$ to obtain all two-point observables.
We observe markedly lower errors compared to the 2D-TNS results from Ref.~\cite{king2024} and a lower error than the quantum annealer when using a sufficiently large $\chi$, even for the longest quench time $t_{a} = 20 {\rm ns}$. 
By solely using BP simple update during the evolution we are able to reach a much larger bond dimension $\chi_{\rm BP}$ than the simulations in Ref.~\cite{king2024} and keep up with the entanglement generated during the quench before truncating the final state at the end. We find this is more favorable than performing aggressive truncation during the evolution \cite{Supplementary}. Moreover, instead of taking samples to calculate observables, we directly measure them with cylindrical MPS message passing, allowing us to avoid statistical measurement errors while maintaining computational efficiency. All $\mathcal{O}(N^{2})$ correlators are computed in $\mathcal{O}(N^{2}\chi^{8})$ time when setting the MPS rank to $R = 2\chi$---which we do here.

In the inset of Figure~\ref{fig:DWaveErrorvsBondDimension}A we also show the anticipated growth of the squared spin glass order parameter \cite{king2023} for a single disorder instance $\langle q^{2} \rangle = \frac{2}{N(N-1)}\sum_{i > j}\langle \sigma^{z}_{i}\sigma^{z}_{j} \rangle^{2}$ with total annealing time $t_a$, computed from the correlators $\langle \sigma^{z}_{i}\sigma^{z}_{j} \rangle$ obtained at the end of the annealing process. 
The results for larger annealing times indicate the glassy, quantum order captured by our cylindrical tensor network ansatz and are in agreement with converged MPS simulations to two significant figures.

\subsection*{Varying the lattice geometry}

\begin{figure*}[t!]
    \centering
    \includegraphics[width =\textwidth]{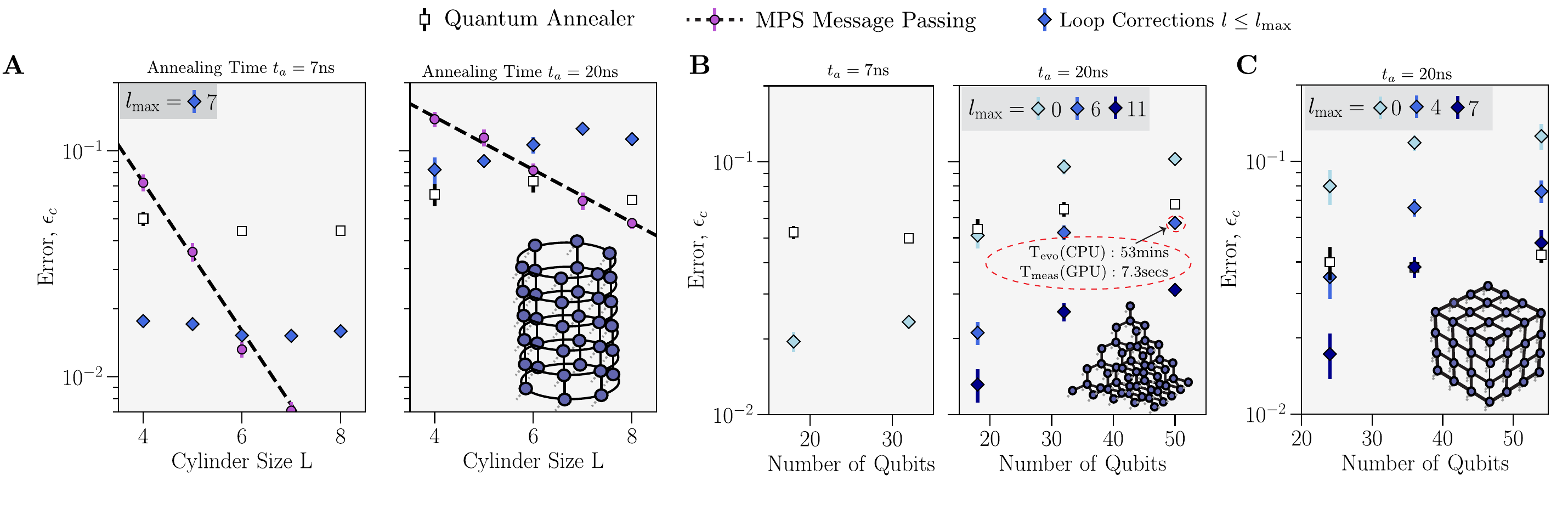}
    \caption{\textbf{Cylindrical, Diamond and Cubic Lattices.}
    Error $\epsilon_{c}$---see Eq.~\ref{Eq:ErrorMetric}---from two- and three-dimensional tensor network (TN) simulations of a glassy quantum annealing protocol for annealing times $t_{a} = 7{\rm ns}$ and $t_{a} = 20{\rm ns}$. We use the same $n = 20$ disorder realizations as in Ref. \cite{king2024}. Error bars correspond to the standard error on the mean. The tensor network is evolved with a Belief Propagation-based evolution protocol with maximum bond dimension $\chi_{\rm BP}$. Expectation values are obtained from the TNS with either cylindrical message passing or loop corrections 
    with a maximum number $l_{\rm max}$ of antiprojectors (see Eq.~\ref{eq:LoopCorrectedSumApprox}). (\textbf{A}) $L \times L$ cylindrical lattice geometry with $\chi_{\rm BP} = 32$ and the final state truncated, via BP, to $\chi = 10$. Message passing is performed with MPS dimension $R = 2\chi$. (\textbf{B-C}) Three-dimensional diamond cubic and dimerized cubic lattice geometries with $\chi_{\rm BP} = 16$ and $\chi_{\rm BP} = 8$ respectively. The final states are not truncated further ($\chi = \chi_{\rm BP}$). The circled data point is annotated with the average clock time for simulating the dynamics ($T_{\rm evo}$) on an Intel Xeon Gold 6244 CPU and for measuring a single $z-z$ correlator ($T_{\rm meas}$) on an Nvidia RTX A6000 GPU.  
    } 
    \label{fig:DWaveErrorvsSystemSize}
\end{figure*}

In Figure~\ref{fig:DWaveErrorvsSystemSize} we consider the cylindrical lattice alongside the diamond cubic and dimerized cubic lattices and compute the error in our simulations as a function of system size.
In the cylindrical case we compare both MPS message passing and loop-corrected contraction schemes to compute the relevant observables. In the three-dimensional cases it is not possible to use MPS message passing efficiently so we compare different levels of loop corrections for computing observables, accounting for loop configurations up to a certain number $l_{\rm max}$ of antiprojectors.

In the cylindrical system we observe an error from MPS message passing which decreases exponentially with the cylinder size. This is caused by the BP approximation being made around the cylinder, although the total error should eventually saturate to a nonzero value commensurate with the bond dimension used for the gate evolution and the boundary MPS rank used to compute observables. 
For both annealing times we see the error converging to a value below that obtained from the quantum annealer. 
For larger cylindrical lattices we find MPS message passing is favorable compared to loop corrections for measuring observables.

In the diamond and cubic lattices we obtain successful results using the loop correction approach, where we find that using larger configuration sizes leads to a significant decrease in the error. In the case of the diamond lattice the error saturates to one noticeably below that of the quantum annealer while on the cubic lattice the error is comparable to the annealer. This is likely because on the $36$ and $54$ qubit dimerized cubic lattices a periodic boundary is present in the $z$-direction which, thanks to the small size of the system, creates a loop of size $3$ in the tensor network. This small loop increases the error in our methods and we anticipate that error should diminish when moving to larger cubic lattices where the smallest loop size will change from $3$ to $4$.

\subsection*{Scalability and Kibble-Zurek Scaling}

\begin{figure*}[t!]
    \centering
    \includegraphics[width =\textwidth]{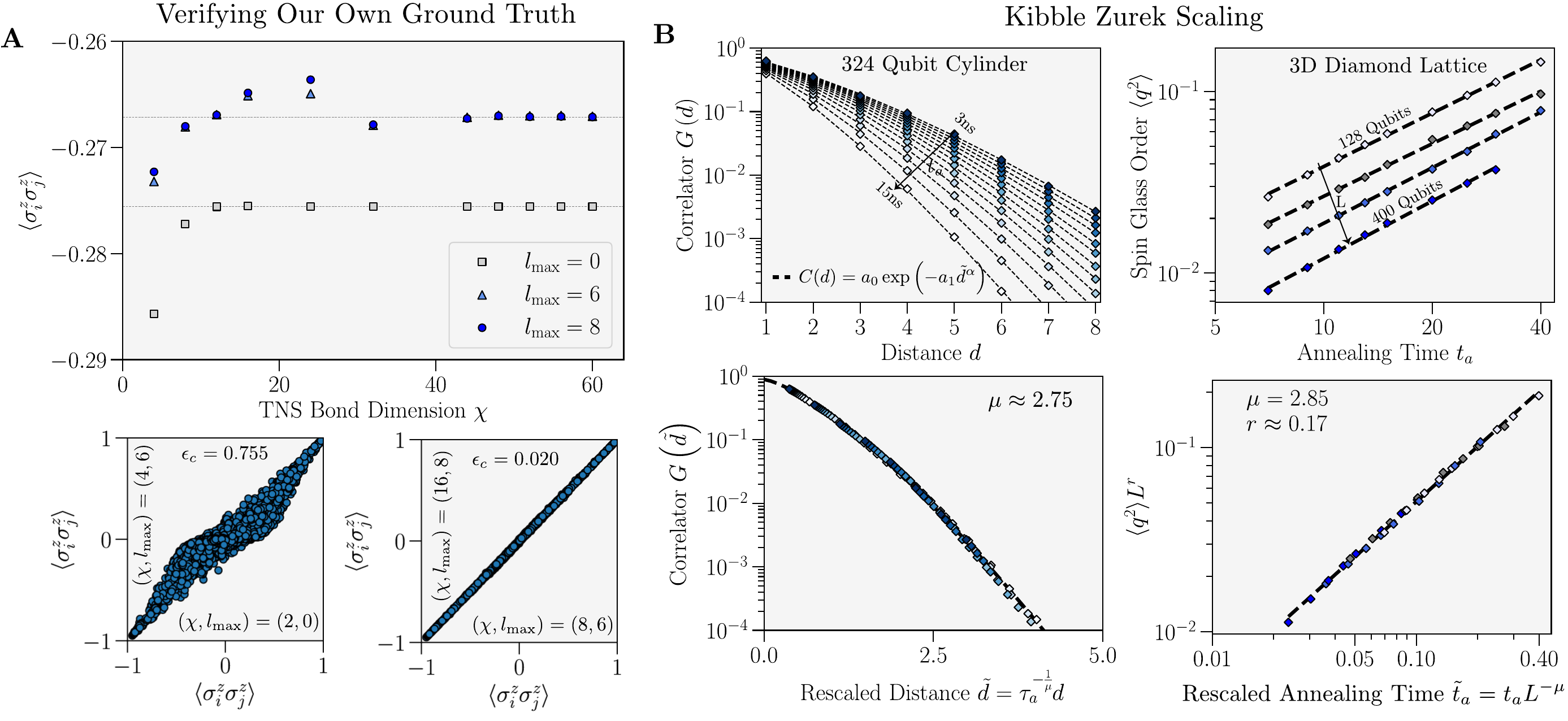}
    \caption{\textbf{Verifying our own ground truth and Kibble-Zurek scaling.} (\textbf{A}) 3D Tensor Network simulation, with bond dimension $\chi$, for an $N = 192$ qubit diamond lattice and annealing time $t_{a} = 7{\rm ns}$. Following evolution, $\langle \sigma^{z}_{i}\sigma^{z}_{j} \rangle$ is calculated via loop corrections of up to size $l_{\rm max}$. Top panel shows an example nearest-neighbor correlator versus $\chi$ for different orders of correction. Bottom panels are scatter plots ($\epsilon_{c}$ is computed via Eq.~\ref{Eq:ErrorMetric}) of all $18,336$ two-point $z-z$ correlations against each other for: (left) $\chi =2, l_{\rm max} = 0$ versus $\chi =4, l_{\rm max} = 6$ and (right) $\chi =8, l_{\rm max} = 6$ versus $\chi =16, l_{\rm max} = 8$. 
    (\textbf{B}) Left panels (top to bottom) show a correlation function collapse on a cylindrical lattice based on  $\langle \sigma^{z}_{i}\sigma^{z}_{j} \rangle$ between vertically aligned spins. The correlation function is extracted via Eq.~\ref{Eq:CorrelationFunction}.
    Distances are rescaled as $\tilde{d} = t_{a}^{-\frac{1}{\mu}}d$ to obtain collapse. Right panels (top to bottom) show spin glass order $\langle q^{2} \rangle$ collapse in a 3D diamond lattice. Collapse is obtained when rescaling the y-axis by $L^{r}$ and the x-axis with $L^{\mu}$ where $L = \sqrt[3]{N}$, with $N$ the number of qubits.
    } 
    \label{fig:DWaveConvergence}
\end{figure*}

Importantly, in all of these simulations, for fixed hyperparameters (e.g.~maximum configuration size $l_{\rm max}$,  MPS message passing rank $R$, and annealing time $t_{a}$), the methods we use to contract the tensor networks in our simulations scale linearly in walltime with the number of qubits (see fig.~\ref{fig:Timing}). As shown in Fig.~\ref{fig:DWaveErrorvsSystemSize}, for the system sizes where we can verify a ground truth and using computational resources that are linear in the system size, the errors in the three-dimensional lattices, at worst, grow only modestly in the system size whereas in the two-dimensional case they are decreasing.

In fig.~\ref{fig:ErrorAnalysis}, on $3$D diamond lattices with up to $900$ qubits, we analyse various error metrics  that do not require ground truth comparison, such as the average gate error and the relative deviation between BP and first order loop corrections. For fixed annealing time, we find these metrics initially increase with system size before quickly saturating to finite values for inputs with $N \geq 100$ qubits. This is consistent with the results in Fig.~\ref{fig:DWaveErrorvsSystemSize} and provides strong evidence for the scalability of our approach.  Moreover, for a fixed annealing time the correlation length, and therefore the bond dimension, are expected to be finite and obey Kibble-Zurek scaling and these results appear to be a reflection of this. As expected, we find the bond dimension required to reach a fixed error increases with the annealing time owing to the increased correlation length.

To provide further evidence of the scalability of our 3D simulations, we show how we can verify our own ground truths via convergence in both bond dimension and loop size with our 3D tensor network approach on a single $192$ qubit diamond input with $t_{a} = 7{\rm ns}$. At this system size it is not feasible to obtain a converged ground truth with an MPS approach due to the prohibitive bond dimensions required \cite{king2024}.

Specifically, in Fig.~\ref{fig:DWaveConvergence}A we show the convergence of a nearest neighbor correlator in both bond dimension $\chi$ and maximum loop size $l_{\rm max}$, leveraging GPUs to significantly lower the loop contraction walltimes at large bond dimension (see fig.~\ref{fig:Timing}). We observe convergence of all orders of loop corrections with increasing $\chi$, with the zeroth ($l_{\rm max} = 0$) and first ($l_{\rm max} = 6$) order loop-corrected values agreeing in the first decimal place whereas the first ($l_{\rm max} = 6$) and second ($l_{\rm max} = 8$) show agreement to the third decimal place. Moreover, we
perform scatter-plot comparisons of all $18,336$ two-point $z-z$ observables computed with different $\chi$ and $l_{\rm max}$. As these values are increased the observables become highly correlated, with the correlation metric $\epsilon_{c}$ (using the larger $\chi$ result as a ground truth) converging to values on the order $\epsilon_{c} \sim 0.02$.

These results all demonstrate the efficacy of a message passing-based structured tensor network approach to simulating the dynamics induced by Hamiltonian Eq.~\ref{Eq:Hamiltonian}.  We leverage the scalability of this approach to demonstrate, in Fig.~\ref{fig:DWaveConvergence}B, a universal collapse of the correlation function and total spin glass order for large cylindrical and diamond lattices respectively as predicted by the theory of non-equilibrium critical phenomena.

In the cylindrical case we compute the correlation function
\begin{equation}
    C(d) = \overline{\vert \langle \sigma^{z}_{i} \sigma^{z}_{j} \rangle - \langle \sigma^{z}_{i} \rangle \langle \sigma^{z}_{j} \rangle \vert }_{{\rm dist}(i,j) = d},
    \label{Eq:CorrelationFunction}
\end{equation}
where the average is taken over ten disorder realizations and ${\rm dist}(i,j)$ is defined as the length of the path (i.e.~the Manhattan distance) between the qubits $i$ and $j$. We consider only non-boundary qubits that lie along the same column in the cylinder. We find that, for a given annealing time and cylinder size, the correlation function can be fit well by a compressed exponential  function  \cite{Trachenko2021, Wu2018} $C(d) \sim a_{0}\exp(-a_{1} d^{\alpha})$ with $\alpha > 1$ over moderate distances ($d < 10$).  We typically observe $\alpha \sim 1.3$.

Upon rescaling the distance as $\tilde{d} = t_{a}^{-\frac{1}{\mu}}d$ where $\mu$ is the Kibble-Zurek exponent we then observe the anticipated collapse of the correlation function \cite{Keesling2019}. We identify \cite{Supplementary} a value of $\mu \approx 2.75$ for the given system, consistent with the very recent Monte-Carlo  based finite-size prediction of Ref.~\cite{Massimo2024} and from the collapse of the Binder cumulant for the three largest cylinders in Ref.~\cite{king2024} which yielded $\mu = 2.67 \pm 0.29$. Further simulations are likely to prove fruitful for determining the scaling of $\mu$ with system size and providing accurate quantification of the uncertainties in $\mu$.

Meanwhile, in the diamond lattices, we extract the Edwards-Anderson, or spin glass, order parameter. A collapse is obtained when rescaling the y-axis by $L^{r}$ and the x-axis with $L^{\mu}$ where $L = \sqrt[3]{N} = \sqrt[3]{\frac{L_{x}L_{y}L_{z}}{4}}$. We set $\mu$ to the value recently obtained for 3D Ising models in Ref. \cite{king2023} and determine a value of $r \approx 0.17$ via a fitting procedure \cite{Supplementary}.

\subsection*{Discussion and outlook}
In this work we have demonstrated that tensor networks, when contracted with message passing-based schemes, can be used to efficiently simulate the complex quantum dynamics induced by the disordered, time-dependent Hamiltonian in Eq.~\ref{Eq:Hamiltonian} on various lattices, including lattices with small loops and in three dimensions. 
Our simulations are scalable in both two and three dimensions, reaching state-of-the-art accuracies with resources that scale roughly linearly in the system size. We are able to reach system sizes on the order of hundreds of qubits, allowing us to observe the universal Kibble-Zurek physics which emerges when driving the system through a phase transition. Moreover, we obtain values for the Kibble-Zurek exponent consistent with recent Monte Carlo-based simulations \cite{Massimo2024}, and the collapse of the Binder cumulant performed in Refs.~\cite{king2024, king2023}.

For the Ising quantum spin glass problem at hand, our classical approach demonstrably outperforms other reported methods, specifically other classical tensor network approaches and machine learning-based approaches \cite{king2024}. In the case of the cylindrical and diamond lattices we are also able to reach errors noticeably lower than the quantum annealing approach employed by the D-Wave Advantage$2$ system \cite{king2024} with a computational scaling that is roughly linear in the system size. In the case of the dimerized cubic lattice, our errors are comparable to the D-Wave system. 

The tensor network methods we use can be directly applied to simulate a wide range of two- and three-dimensional systems. Specifically, we anticipate significant success in their application to tasks such as solving optimization problems via simulated quantum annealing and simulating the finite-temperature physics of three-dimensional spin-ice models \cite{PhysRevLett.115.077202}. 

Crucially, we also expect the efficiency and scalability of higher-dimensional tensor network methods will continue to improve. 
New approaches leveraging the flexibility of belief propagation and its extensions, the continued growth of processing power of classical computers, and the parallelism latent in these methods will continue to push the boundaries of classical methods for quantum simulation.

\newpage




\clearpage 

%


\section*{Acknowledgments}
We thank Steve White and Matthias Troyer for helpful discussions and comments.
\paragraph*{Funding:}
The authors are grateful for ongoing support through the Flatiron Institute, a division of the Simons Foundation. 
D.S. was supported by AFOSR: Grant FA9550-21-1-0236.
\paragraph*{Author contributions:}
Conceptualization: JT. Methodology: JT, MF, MS. Investigation: JT, AM, DS. Visualization: JT, AM. Project Administration: JT, DS. Supervision: JT, DS. Writing - original draft: JT, DS. Writing - review and editing: JT, MS, MF, DS, AM.
\paragraph*{Competing interests:}
Authors declare that they have no competing interests.
\paragraph*{Data and materials availability:}
 Software for performing the simulations in this paper is available at \textsc{TensorNetworkQuantumSimulator.jl} \cite{TensorNetworkQuantumSimulator}. The data presented in this paper are archived at \cite{DryadData} and can be used to directly reproduce all of the figures in the main text and supplementary material. No new materials were generated for this study.


\subsection*{Supplementary materials}
Supplementary Text \\
Figs. S1 to S5\\
References \textit{(54-\arabic{enumiv})}\\ %


\newpage


\renewcommand{\thefigure}{S\arabic{figure}}
\renewcommand{\thetable}{S\arabic{table}}
\renewcommand{\theequation}{S\arabic{equation}}
\renewcommand{\thepage}{S\arabic{page}}
\setcounter{figure}{0}
\setcounter{table}{0}
\setcounter{equation}{0}
\setcounter{page}{1} 


\begin{center}
\section*{Supplementary Materials for\\ \scititle}

Joseph~Tindall$^{1\ast}$,
	Antonio~Francesco~Mello$^{1, 2}$,
	Matthew~Fishman$^{1}$, \and
    E.~Miles~Stoudenmire$^{1}$,
    Dries~Sels$^{1,3}$
\end{center}

\subsubsection*{This PDF file includes:}
Materials and Methods \\
Supplementary Text \\
Figs. S1 to S5\\
References \textit{(54-\arabic{enumiv})}\\ %

\newpage

\subsection*{Materials and Methods}
Here we outline our tensor network-based simulation approach in detail. We model the wavefunction under the time-dependent Hamiltonian $H(s)$ in Eq.~\ref{Eq:Hamiltonian} as a tensor network of low-rank tensors whose structure mimics that of the underlying lattice geometry - i.e.~for every pair of neighboring points in the lattice there exists a virtual edge or bond between the corresponding tensors in the tensor network. In the case of the dimerized cubic lattice, we pair the dimers together into a single tensor with two physical legs, effectively realizing a regular cubic tensor network.
This removes the inherent loops that form from couplings in the $z$-direction between the spins in the dimers and thus make the structure more ``tree-like''.

\subsubsection*{Belief Propagation and Loop Corrections}
\begin{figure*}[t!]
    \centering
    \includegraphics[width =\textwidth]{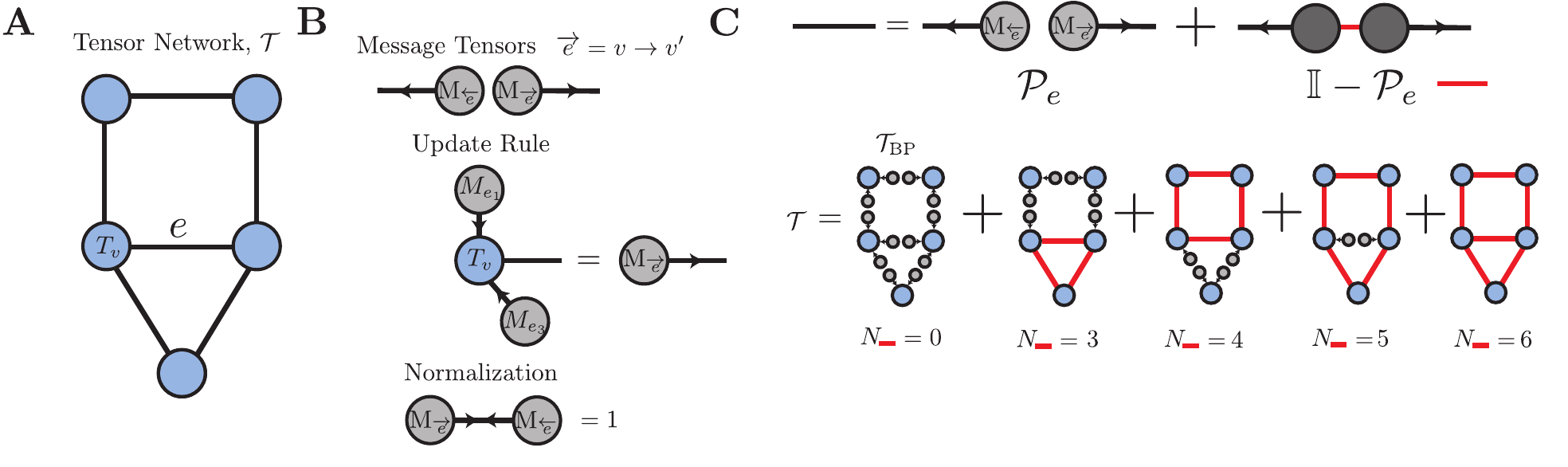}
    \caption{
    (\textbf{A}) Tensor network on a simple $5$-vertex graph. The vertices may represent individual tensors or groups of tensors.
    (\textbf{B}) Belief propagation involves defining message tensors along both directions of the edges of the norm network with the indices of the message tensors matching the virtual indices of the bra and ket tensors along that edge. An update rule is defined for every message tensor as the product of the messages incident to the vertex the message is leaving and the tensors on the vertex. This rule is iterated until all messages are converged, with a normalization rule imposed for numerical stability. (\textbf{C}) The outer product of the two message tensors on an edge of $\mathcal{T}$ form a projector $\mathcal{P}_{e}$ to a rank-1 subspace. The full contraction of the network can then be written as a sum over all $2^{N_{\rm edges}}$ configurations involving either a projector or its antiprojector $\mathbb{I} - \mathcal{P}_{e}$. Only the nonzero terms are shown: any configuration where there exists a vertex where only one antiprojector is incident is zero. The contraction of the network can then be approximated by summing configurations up to a certain number of antiprojectors, with the term with zero antiprojectors corresponding to the uncorrected belief propagation result.}
    \label{fig:BeliefPropagation}
\end{figure*}

In this work we use belief propagation \cite{Sahu2022, Alkabetz2021, tindallbeliefpropagation} and extensions thereof to evolve the tensor network state $\vert \psi \rangle$ and compute observables. 
For evolving the state we use regular (non-corrected) belief propagation due to its efficiency and simplicity, which is equivalent to performing gate evolution with the standard simple update method \cite{Jiang2008, Orus2019, Alkabetz2021, tindallbeliefpropagation} but reformulated in a more generalizable framework. One method we use for computing observables is loop-corrected belief propagation \cite{evenbly2024} to improve the accuracy of our results beyond the belief propagation approximation.
Both tasks rely on computing ``message tensors'' which form rank-one projectors to the virtual basis defined on the edges of a tensor network $\mathcal{T}$ whose vertices represent either individual tensors or groups of tensors. 
In fig.~\ref{fig:BeliefPropagation} we illustrate the belief propagation algorithm for an example network $\mathcal{T}$ where the vertices of the network $\mathcal{T}_{v}$ may represent individual tensors or collections of tensors. In the case of a norm network, defined as $\mathcal{T} = \langle \psi \vert \psi \rangle$, the vertices represent uncontracted pairs of bra and ket tensors from $\vert \psi \rangle$, that is $\mathcal{T}_{v} = \psi_{v} \psi_{v}^{*}$. Message tensors are defined along the edges of the norm network whose indices match the virtual indices connecting the corresponding bra and ket tensors on each end of the edge. A self-consistent update rule is defined for each message tensor as
\begin{equation}
    M_{v \rightarrow v'} = \left(\prod_{v'' \in \{{\rm neighbors}(v) / v'\}}M_{v'' \rightarrow v}\right)\mathcal{T}_{v}
\end{equation}
and this rule is iterated until convergence of the message tensors. 
Imposing the normalization condition \newline $M_{v \rightarrow v'} \cdot M_{v' \rightarrow v} = 1$ is helpful both for numerical stability and for later computations with the message tensors. 
For $\mathcal{T} = \langle \psi \vert \psi \rangle$, a single iteration that updates every message scales as $\mathcal{O}(N \chi^{z+1})$ in time where $N$ is the number of tensors in $\vert \psi \rangle$, $z$ is the maximum coordination of the vertices of the graph, and $\chi$ is the maximum dimension of the virtual indices in $\vert \psi \rangle$. Convergence of the message tensors is typically exponentially fast in the number of iterations.

The resulting message tensors can be used in a variety of ways. 
For instance, they can be used to perform a truncation following a two-site update with a gate $\hat{G}$ on a neighboring pair of sites in $\vert \psi \rangle$ \cite{tindallbeliefpropagation}. 
This is the method we adopt in our time-evolution protocol due to its numerical efficiency and simplicity. 
Message tensors can also be used to compute local and non-local observables $\langle O \rangle$. 
At the simplest, but most approximate, level one just places the message tensors obtained from $\langle \psi \vert \psi \rangle$ incident to the region of support of $O$, inserts $O$, and contracts the tensors to obtain the expected value of $O$. 
If the network is unnormalized the result can be divided by the same contraction but without $O$ inserted.
Alternatively, belief propagation can be run  directly on the network $\langle \psi \vert O \vert  \psi \rangle$, grouping together the local bra, ket, and operator tensors, and then the resulting message tensors can be used to approximate the contraction of the whole network via the formula
\begin{equation}
    \langle \psi \vert O \vert \psi \rangle = \frac{\prod_{v \in {\rm verts}}\left(\left(\psi_{v}O_{v}\psi^{*}_{v}\right)\prod_{v' \in {\rm  neighbors}(v)}M_{v' \rightarrow v}\right)}{\prod_{e \in {\rm edges}}M_{e} \cdot M_{{\rm reverse}(e)}},
    \label{Eq:BPEquation}
\end{equation}
which is more appropriate in the case of non-local observables with a large region of support. If the network is unnormalized an equivalent contraction should be done on the network $\langle \psi \vert \psi \rangle$ and used as a divisor.

Both of the above approaches, however, do not account for the loops in the lattice and thus make a fairly severe approximation for correlated states and tensor networks with a larger density of small loops. Nonetheless, the messages obtained from belief propagation can be used as a basis from which to expand the contraction of a tensor network as a multi-dimensional sum which contains ``loop-corrected'' terms \cite{evenbly2024}.

In the loop-corrected approach the contraction of the target tensor network $\mathcal{T}$ (with tensors or tensor groups $T_{1}, T_{2}, ... T_{N}$) upon which message tensors have been computed is written as a sum over $2^{N_{\rm edges}}$ configurations involving the placement of either a rank-1 projector $\mathcal{P}_{e} = M_{\overleftarrow{e}} \otimes M_{\overrightarrow{e}}$ or an antiprojector $\mathbb{I} - \mathcal{P}_{e}$ on each of the edges $e$ of the network. Specifically, letting $\mathcal{S}$ denote some subset of the set of edges $E$, we can define the term
\begin{equation}
    W_{\mathcal{S}} = \left(\prod_{v=1}^{L}T_{i} \right)\left(\prod_{e \in \mathcal{S}}(\mathbb{I} - \mathcal{P}_{e})\right) \left(\prod_{e \in E / \mathcal{S}}\mathcal{P}_{e}\right),
\end{equation}
involving the placement of $\vert \mathcal{S} \vert$ antiprojectors in the specified configuration and $N_{\rm edges} - \vert \mathcal{S} \vert$ projectors on the remaining edges of the network. 

Then, the full contraction of the network is given by the sum
\begin{equation}
    \mathcal{T} = \sum_{i=0}^{N_{\rm edges}}\sum_{\mathcal{S} \in \binom{E}{i}}W_{\mathcal{S}},
    \label{eq:LoopCorrectedSum}
\end{equation}
where $\binom{E}{i}$ denotes the set of all subsets of the edges of the tensor network of size $i$. It is helpful to note that, due to the fixed point condition on the converged message tensors (see Eq.~\ref{Eq:BPEquation}), the term $W_{\mathcal{S}}$ is zero if there exists any vertex $v$ for which only one antiprojector is incident. Since the magnitude of a single term $W_{\mathcal{S}}$ should generically decay exponentially with $\vert S \vert$, an efficient, but potentially highly accurate, approximation for the contraction of the network can then be found by summing up all nonzero configurations with some maximum threshold number of $l_{\rm max} \ll N_{\rm edges}$ antiprojectors, that is
\begin{equation}
    \mathcal{T} \approx \sum_{i=0}^{l_{\rm max}}\sum_{\mathcal{S} \in \binom{E}{i}}W_{\mathcal{S}} \ .
    \label{eq:LoopCorrectedSumApprox}
\end{equation}

Supplementary Figure \ref{fig:BeliefPropagation}C shows the loop correction procedure for an example tensor network, illustrating all the nonzero $W_{s}$ terms that form the sum in Eq.~\ref{eq:LoopCorrectedSum}. This is the approach we take for computing ``loop-corrected'' observables such as 
\begin{equation}
    \langle \sigma^{z}_{i} \sigma^{z}_{j} \rangle = \frac{\langle \psi \vert \sigma^{z}_{i} \sigma^{z}_{j} \vert \psi \rangle}{\langle \psi \vert \psi \rangle},
    \label{Eq:Ratio}
\end{equation}
for any $i,j$ from the tensor networks $\vert \psi \rangle$ we obtain in this work. While $\langle \sigma^{z}_{i} \sigma^{z}_{j} \rangle$ can be computed via separate loop-corrected contractions of the numerator and denominator in Eq.~\ref{Eq:Ratio}, we find that finding BP fixed points of tensor networks such as $\langle \psi \vert \sigma^{z}_{i} \sigma^{z}_{j} \vert \psi \rangle$ which are not positive-definite (i.e. they are not of the form $O = XX^{\dagger}$) can be difficult. Thus it is beneficial to first recast $\langle \sigma^{z}_{i} \sigma^{z}_{j} \rangle$ in terms of a combination of observables which are all positive-definite:
\begin{equation}
    \langle \sigma^{z}_{i} \sigma^{z}_{j} \rangle = \frac{\langle \psi \vert P^{\uparrow}_{i}P^{\uparrow}_{j} \vert \psi \rangle + \langle \psi \vert P^{\downarrow}_{i}P^{\downarrow}_{j} \vert \psi \rangle - \langle \psi \vert P^{\uparrow}_{i}P^{\downarrow}_{j} \vert \psi \rangle- \langle \psi \vert P^{\downarrow}_{i}P^{\uparrow}_{j} \vert \psi \rangle}{\langle \psi \vert P^{\uparrow}_{i}P^{\uparrow}_{j} \vert \psi \rangle + \langle \psi \vert P^{\downarrow}_{i}P^{\downarrow}_{j} \vert \psi \rangle + \langle \psi \vert P^{\uparrow}_{i}P^{\downarrow}_{j} \vert \psi \rangle + \langle \psi \vert P^{\downarrow}_{i}P^{\uparrow}_{j} \vert \psi \rangle} = \frac{\langle \psi \vert P^{\uparrow}_{i}P^{\uparrow}_{j} \vert \psi \rangle - \langle \psi \vert P^{\downarrow}_{i}P^{\uparrow}_{j} \vert \psi \rangle}{\langle \psi \vert P^{\uparrow}_{i}P^{\uparrow}_{j} \vert \psi \rangle + \langle \psi \vert P^{\downarrow}_{i}P^{\uparrow}_{j} \vert \psi \rangle}
    \label{Eq:PosDefRecact}
\end{equation}
where $P_{i}^{\uparrow}$ and $P_{i}^{\downarrow}$ are the projectors to spin up and spin down respectively on the given site and the last equality follows only due to the $\mathbb{Z}_{2}$ symmetry of the wavefunction ((i.e $\left(\prod_{i}\sigma^{x}_{i} \right)\vert \psi \rangle = \vert \psi \rangle$) which we find to be well preserved during our TN-based evolution despite no explicit encoding of symmetries into the tensors. We note that such symmetries could be encoded directly at the level of the individual tensors, and potentially improve the efficiency of our calculations---something we leave for future work.
\par We thus compute $\langle \sigma^{z}_{i} \sigma^{z}_{j} \rangle$ by using loop corrections to contract the two positive definite tensor networks $\langle \psi \vert P^{\uparrow}_{i}P^{\uparrow}_{j} \vert \psi \rangle$ and $\langle \psi \vert P^{\downarrow}_{i}P^{\uparrow}_{j} \vert \psi \rangle$ and substituting into Eq.~\ref{Eq:PosDefRecact}. In the case the $\mathbb{Z}_{2}$ symmetry is not present, loop corrections can be used to approximate the contraction of the four networks from Eq.~\ref{Eq:PosDefRecact} instead. For each network, we group together tensors associated with each site and then run belief propagation over the resulting partitioned network (whose vertices represent the corresponding groups of tensors) to obtain converged message tensors.
We use a well-established graphical loop counting algorithm \cite{Johnson1975, Fairbanks2021} to enumerate short loops in our networks and then compose the loops together to efficiently enumerate all configurations for which $W_{S}$ is nonzero and the number of antiprojectors is less than some amount $l_{\rm max}$. 
We contract these configurations, something which can be done in a fully parallelizable manner, and add them up to form a loop-corrected approximation for  $\langle \sigma^{z}_{i} \sigma^{z}_{j} \rangle$. In the main text we show our results using different values of $l_{\rm max}$. The computational scaling of this approach depends on $l_{\rm max}$ and the coordination number $z$ of the underlying lattice. Evaluating $\mathcal{T} = \langle \psi \vert O \vert \psi \rangle$ (where $O$ is a product of local spin-1/2 operators and tensors sharing site indices are grouped together) with $l_{\rm max} = 0$, for instance, can be achieved in $\mathcal{O}(N\chi^{z+1})$ time. Meanwhile, the next order correction will involve $l_{\rm max}$ being set to the size of the shortest loop in the tensor network and its evaluation scales in time linearly in the loop length and as $\mathcal{O}(N \chi^{z+2}) + \mathcal{O}(N\chi^{6})$ with the bond dimension and system size---although we remark that if the loop is sufficiently large, methods like randomized SVD or Krylov solvers can be used to accelerate the scaling to $\mathcal{O}(mN \chi^{z+1})$ where $m$ is the number of dominant eigenmodes targeted in the decomposition \cite{Pippan2010}.  

While in this work we run loop-corrected belief propagation to obtain messages and contract the configurations in each network separately, Eq.~\ref{eq:LoopCorrectedSum} holds for arbitrary message tensors provided the normalization condition $M_{e} \cdot M_{\rm reverse(e)} = 1$ holds. 
More efficient ways to obtain multiple observables at once via loop corrections could be achieved by using a consistent set of message tensors (say those from $\langle \psi \vert \psi \rangle$ or $\langle \psi \vert O' \vert \psi \rangle$ where $O'$ is some subset of the local operators that make up the full operator $O$) across the expansion of the different observables $\{\langle \psi \vert O_{1} \vert \psi \rangle, \langle \psi \vert O_{2} \vert \psi \rangle, ... \}$ that need to be measured. 
Therefore, we could cache certain loop contractions which appear multiple times in the expansion of various observables.
Care would need to be taken, however, to ensure the message tensors used still form a good basis for each of the individual networks $\langle \psi \vert O_{i} \vert \psi \rangle$, which is likely to be the case only if the support of the relevant observable is small. 
 
\subsubsection*{Cylindrical MPS Message Passing: }
\begin{figure*}[t!]
    \centering
    \includegraphics[width =\textwidth]{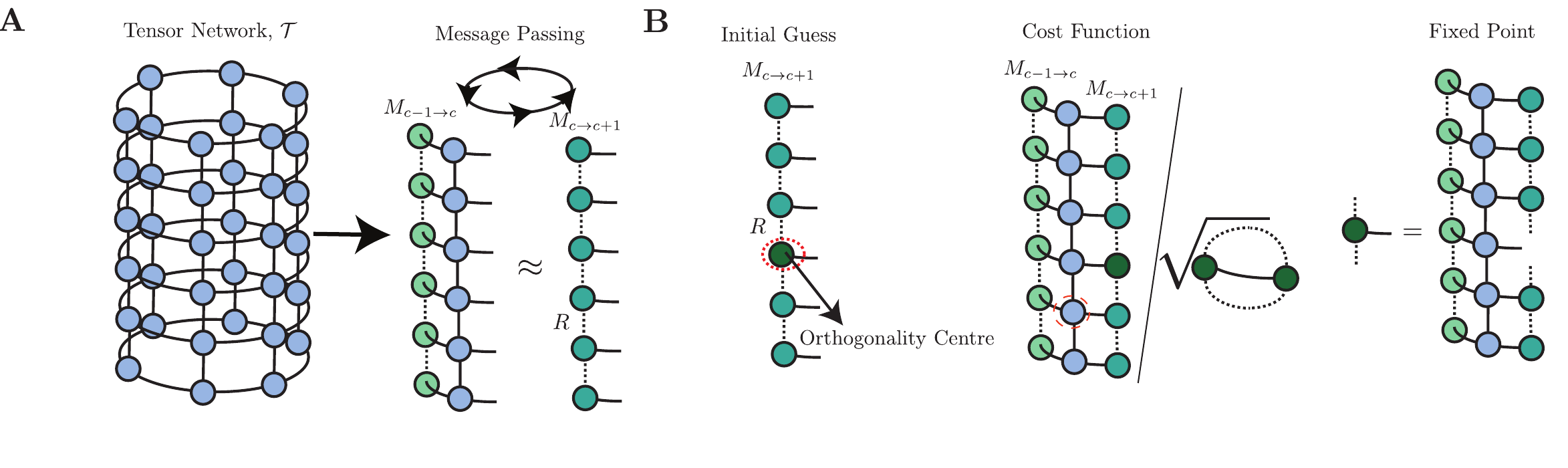}
    \caption{ (\textbf{A}) Matrix product state message passing on a cylindrical tensor network. The tensors in the network could represent single tensors or groups of tensors.
    Matrix product states are repeatedly passed  around the cylinder a finite number of times until they converge. (\textbf{B}) One-site optimization procedure for the matrix product state message passing. An initial matrix product state is used as a guess for the outgoing MPS from the product of an incoming matrix product state and the column. The tensor on the orthogonality center of the outgoing MPS can be updated with the derivative of the corresponding MPS-MPO-MPS network with respect to that tensor. By repeatedly sweeping up and down the MPS, moving the orthogonality center and caching contractions of the MPS-MPO-MPS structure, the sites of the outgoing MPS can be efficiently optimized one-by-one until convergence of the cost function.}
    \label{fig:CylindricalMessagePassing}
\end{figure*}

For the case of a cylindrical tensor network we find, for sufficiently large cylinder sizes, that cylindrical message passing is a more effective method than loop corrections for computing two-point observables in the network. This method can still be viewed as a variant of belief propagation (see also Ref. \cite{Guo2023} which introduces a related hybridization of the boundary MPS and BP algorithms) where the network has been partitioned via its columns into a ring and MPS messages leaving a column are iteratively updated as the truncated product of the incoming message to that column and the column itself (see fig.~\ref{fig:CylindricalMessagePassing}A). Upon convergence the two MPS messages incident to a column form an approximation for the derivative of the network with respect to that column. The degree of the approximation depends on two factors: the maximum virtual bond dimension allowed in the MPS message tensors and the circumference of the cylinder. The latter is an error that occurs due to the periodic boundary in the system, with the MPS messages effectively ignoring the correlations which flow around the circumference of the cylinder. This error will generically decay exponentially in the circumference of the cylinder  (see Fig.~\ref{fig:DWaveErrorvsSystemSize} of the main text), making the method highly effective for large cylinders.

To compute a MPS message update, we take the outgoing message $\vert M_{c \rightarrow c+1} \rangle$ from a column of tensors $\mathcal{T}_{c}$ to be the one with a fixed virtual dimension $R$ that maximizes the cost function
\begin{equation}
    C = \frac{\langle M_{c -1 \rightarrow c} \vert \mathcal{T}_{c} \vert M_{c \rightarrow c + 1} \rangle}{\sqrt{\langle M_{c \rightarrow c + 1} \vert M_{c \rightarrow c + 1} \rangle}}.
\end{equation}
Assuming the incoming message is fixed, we set the orthogonality center of $\vert M_{c \rightarrow c+1} \rangle$ to the tensor on a given site $s$ and then differentiate the cost function with respect to that tensor. An update of that tensor that maximizes the cost function should then be equal to the contraction of $\langle M_{c -1 \rightarrow c} \vert \mathcal{T}_{c} \vert M_{c \rightarrow c + 1} \rangle$ with the tensor removed (see fig.~\ref{fig:CylindricalMessagePassing}B). The fitting can thus be done by forming an initial guess for $\vert M_{c \rightarrow c+1} \rangle$, iteratively moving the orthogonality center up and down the MPS, and replacing the tensor with the derivative of $\langle M_{c -1 \rightarrow c} \vert \mathcal{T}_{c} \vert M_{c \rightarrow c + 1} \rangle$ with respect to that tensor. The cost function should then converge and the next MPS can be updated until they all converge. For the cylindrical tensor network $\mathcal{T} = \langle \psi \vert O \vert \psi \rangle$, where $O$ is a product of local spin-1/2 operators, identifying the optimal contract order of the tensors leads to a computational scaling in time of $\mathcal{O}(R^{3} \chi^{5})$ if $R \leq \chi$ and $\mathcal{O}(R^{2} \chi^{6}) + \mathcal{O}(R^{3}\chi^{4})$ if $R > \chi$. In this work we set $R = 2\chi$ for our desired accuracy, giving a scaling of $\mathcal{O}(N\chi^{8})$. This approach generalizes immediately to any tensor network $\mathcal{T}$ which forms an open-boundary or half-periodic planar lattice, with the scaling increasing with the coordination number $z$ of the lattice.  

This MPS message passing method can be exploited further to efficiently compute multi-point correlators in the system and we do so in this work. Specifically, consider the goal of computing the set of $N$ two-point correlators $\{\langle \sigma^{z}_{i}\sigma^{z}_{1} \rangle, \langle \sigma^{z}_{i}\sigma^{z}_{2} \rangle, \langle \sigma^{z}_{i}\sigma^{z}_{3} \rangle \hdots \langle \sigma^{z}_{i}\sigma^{z}_{N} \rangle \}$ between site $i$ and the remaining sites in the lattice. By performing our correlated MPS message passing on the positive-definite network $\langle \psi \vert P^{\uparrow}_{i} \vert \psi \rangle$, where $P^{\uparrow}_{i}$ is the spin up projector on site $i$, and then systematically zig-zagging through the columns with the incident MPS messages inserted, we can extract all the one-site reduced density matrices, conditioned on the projection of site $i$ to spin up. In the system at hand the wavefunction is $\mathbb{Z}_{2}$ symmetric (i.e $\left(\prod_{i}\sigma^{x}_{i} \right)\vert \psi \rangle = \vert \psi \rangle$) and so this information is all that is required to compute the set $\{\langle \sigma^{z}_{i}\sigma^{z}_{1} \rangle, \langle \sigma^{z}_{i}\sigma^{z}_{2} \rangle, \langle \sigma^{z}_{i}\sigma^{z}_{3} \rangle \hdots \langle \sigma^{z}_{i}\sigma^{z}_{N} \rangle \}$ in linear time. This method can then be reapplied for each site $i$ to compute all two-point correlators in $\mathcal{O}(N^{2})$ time. This approach can be generalised to cases without the $\mathbb{Z}_{2}$ symmetry by insertion of the operator $\sigma^{z}_{i}$ instead of the projector.  

\subsubsection*{Gate evolution}
The tensor network at renormalized time $s = \frac{t}{t_{a}}$ is evolved by a discrete time step $\delta t$ via the application of a series of one- and two-site gates which stem from a 2nd order Trotter decomposition of the full propagator
\begin{equation}
    U(s)= \exp(-{\rm i}H(s) \delta t) \approx U_{\rm X}(s)U_{\rm ZZ}(s)U_{\rm X}(s)
\end{equation}
with 
\begin{equation}
    U_{\rm X}(s) = \prod_{i=1}^{L} \exp \left(-{\rm i}\frac{\delta t}{2}\Gamma(s)  \sigma^{x}_{i} \right)
    \label{Eq:OneSiteGates}
\end{equation}
and 
\begin{equation}
    U_{\rm ZZ}(s) = \prod_{\langle i,j \rangle} \exp(-{\rm i}\delta t\mathcal{J}(s)\sigma^{z}_{i}\sigma^{z}_{j}).
    \label{Eq:TwoSiteGates}
\end{equation}
The time step $\delta t = 0.01$ns we use is enough for the Trotter error to remain sufficiently low for the accuracy needed in this work.

The one-site gates in Eq.~\ref{Eq:OneSiteGates} can be applied directly to the tensor network without any loss of fidelity or change in the message tensors. Meanwhile, all the two site gates in Eq.~\ref{Eq:TwoSiteGates} act on neighboring tensors in the network and thus we can efficiently use belief propagation-computed environments (see fig.~\ref{fig:BeliefPropagation}B for an illustration on how to compute these environments) to truncate the virtual bond between the corresponding neighboring tensors upon applying the gate. This is equivalent to applying the gate using the standard simple update algorithm \cite{Vidal2003, Vidal2004, Jiang2008, Alkabetz2021, tindallbeliefpropagation}. 
The total time complexity for the simulation of the dynamics is thus $\mathcal{O}\left(N \chi^{z+1} \frac{t_{a}}{\delta t} \right)$ where $N$ is the number of qubits, $\chi$ is the maximum bond dimension allowed in the simulation, and $z$ is the maximum coordination number of the vertices in the underlying lattice ($4$ for cylindrical, $4$ for diamond cubic, and $6$ for the dimerized cubic). To initialize the system in the ground state of $H(0)$ we run imaginary time evolution with the same simple update procedure as we do for the dynamics. As the ground state has only very small correlations ($\mathcal{J}(0) \gg \Gamma(0)$) this is sufficient to obtain a highly accurate initial state.

The final truncation that we perform in the cylindrical lattice case before measuring observables is done via belief propagation \cite{tindallbeliefpropagation}. Note that to obtain the highest accuracy truncation under the BP approximation, one can re-run belief propagation to obtain converged message tensors and use those to perform the truncation, though for small truncation errors and Trotter step sizes the final messages obtained from the evolution may be sufficiently converged already.

\subsection*{Supplementary Text}
\subsubsection*{System size scaling analysis for three-dimensional diamond simulations}

\begin{figure*}[t!]
    \centering
    \includegraphics[width =\columnwidth]{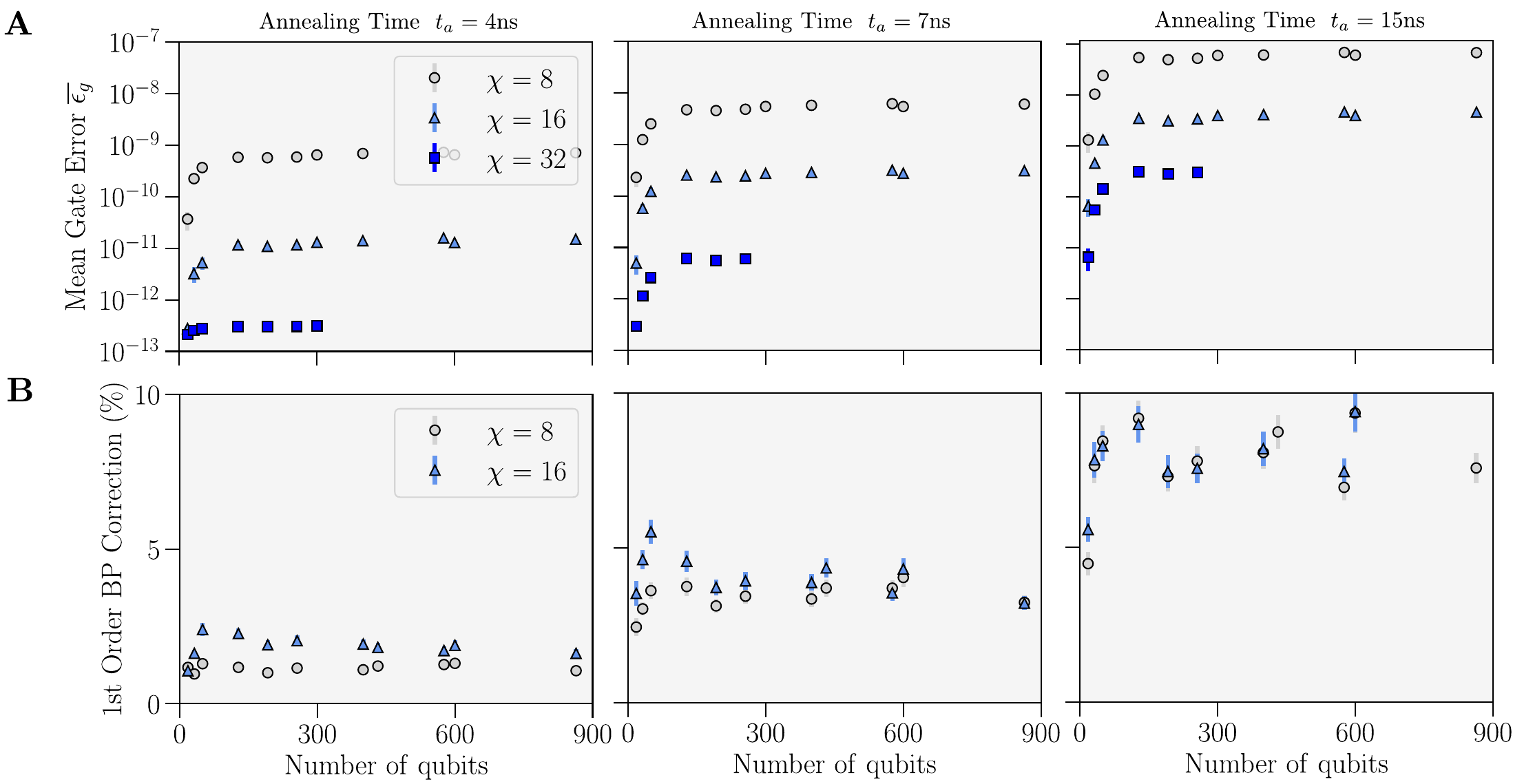}
    \caption{Error metrics for tensor network simulations of 3D diamond inputs across a range of increasing system sizes. Each column shows a different annealing time. (\textbf{A}) Average two-qubit gate error (see Eq.~\ref{Eq:GateError}) versus system size and for various bond dimensions. (\textbf{B}) Averaged relative difference (as a percentage) (see Eq.~\ref{Eq:BPCorrectionPercentage}) between belief propagation ($l_{\rm max} = 0$)  and first order ($l_{\rm max} = 6$) loop corrections for third nearest neighbor correlators. Given the fast saturation of the results with system size, we choose not to run the $\chi=32$ results for larger system sizes for environmental reasons, though we emphasize that the calculations scale only linearly with system size for a given bond dimension}.
    \label{fig:ErrorAnalysis}
\end{figure*} 

Here we provide extensive data and analysis for tensor network simulations on diamond lattices ranging in size from $18$ qubits ($L_{x} \times L_{y} \times L_{z} = 3 \times 3 \times 8$) to $900$ qubits ($L_{x} \times L_{y} \times L_{z} = 12 \times 12 \times 24$). Due to the linear scaling with system size of our simulations for a fixed annealing time and bond dimension, such system sizes are possible with moderate computational resources. Moreover, there are several metrics we can efficiently compute that attest to the ``quality" of the simulation. 
\par The first of these is the average gate error $\overline{\epsilon_{g}}$ which we compute following the time evolution of the tensor network via
\begin{equation}
    \overline{\epsilon_{g}} = \left(1 - \prod_{g =1}^{n} (1 - \epsilon_{g})\right)^{\frac{1}{n}}
    \label{Eq:GateError}
\end{equation}
where the product is over all $n$ two-qubit gates applied to the tensor network during the simulation and $\epsilon_{g}$ is defined as the individual gate error 
\begin{equation}
    \epsilon_{g} = \sum_{i = \chi + 1}^{\chi'}\sigma_{i}^{2},
\end{equation}
i.e. the sum of the square of the discarded singular values during the SVD performed following the application of a gate $\hat{G}$ (where the normalization $\sum_{i=1}^{\chi'}\sigma_{i}^{2} = 1$ is enforced) to the tensor network representation of the state $\vert \psi_{t - \delta t} \rangle$. On a tree tensor network, $\epsilon_{g}$ is precisely the individual gate error \cite{Yiqing2020}, $1 - \vert \langle \psi_{t} \vert \hat{G} \vert \psi_{t - \delta t} \rangle \vert^{2}$ while on a loopy network it is a good approximation \cite{ lee2025scalablesimulationrandomquantum, rudolph2025simulatingsamplingquantumcircuits}.

\begin{figure*}[t!]
    \centering
    \includegraphics[width =\columnwidth]{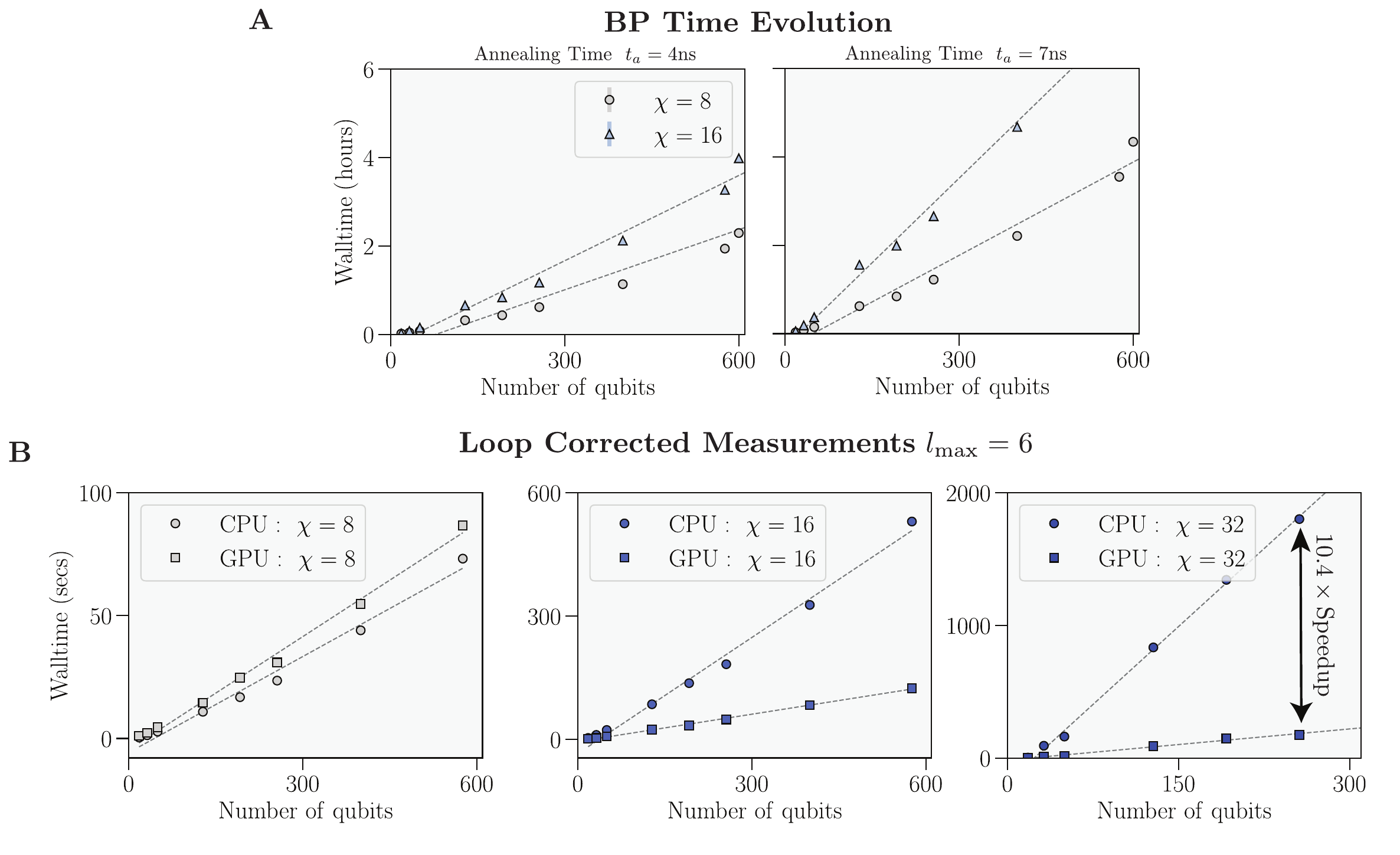}
    \caption{Walltimes for tensor network simulations of 3D diamond inputs across a range of increasing system sizes. Dotted lines represent linear fits. (\textbf{A}) Two different annealing times (left and right panel) and CPU walltime for BP-based time evolution for simulating the time-evolution ($s= 0$ to $s=0.6$). Walltimes are based on multi-threaded simulations using $8$ cores of a single Intel Icelake CPU and averaged over $5$ disorder instances.  (\textbf{B})  Walltimes for measuring a single randomly chosen $z-z$ correlator from 3D tensor networks of various bond dimensions (left, center, and right panel). In all cases $32$-bit floating point precision is used and the tensor network is that obtained with BP-based simple update following an annealing protocol with $t_{a} = 7{\rm ns}$. CPU walltime is based on a multithreaded simulation using $8$ cores of a single Intel Xeon Gold $6244$ CPU. GPU walltime is based on a simulation using an Nvidia RTX A6000 GPU. For large bond dimensions we measure a tenfold speedup when using the GPU versus the CPU.
    }
    \label{fig:Timing}
\end{figure*} 

\par Our second metric is based on the strength of the first order loop corrections in the tensor network following evolution. Specifically we take a random sample of $N$ two-point correlators $\langle \sigma^{z}_{i}\sigma^{z}_{j} \rangle$ where the pair of sites $i$ and $j$  are separated by a fixed path length $d$ and compute
\begin{equation}
    \epsilon_{\rm BP}(d) = 100 \cdot \frac{\sum_{(i,j)}\vert \langle \sigma^{z}_{i}\sigma^{z}_{j} \rangle_{l_{\rm max} = 6} - \langle \sigma^{z}_{i}\sigma^{z}_{j} \rangle_{l_{\rm max} = 0} \vert}{\sum_{(i,j)} \vert \langle \sigma^{z}_{i}\sigma^{z}_{j} \rangle_{l_{\rm max} = 6} \vert}, \qquad {\rm dist}(i,j) = d
    \label{Eq:BPCorrectionPercentage}
\end{equation}
which is a relative measure (as a percentage) of how much the smallest loops in the lattice affect correlators at the given distance $d$.
\par In fig.~\ref{fig:ErrorAnalysis}A we show how the average gate error scales with system size in 3D diamond  lattices for several annealing times ($t_{a} \in \{4 {\rm ns},7{\rm ns},15{\rm ns}\}$) and bond dimensions ($\chi \in \{8,16,32\}$). In all cases increasing the bond dimension leads to systematically lower gate errors while increasing the annealing time leads to larger gate errors, which is also observed for the cylindrical lattice in fig.~\ref{fig:DWaveErrorvsBondDimension} of the main text.
Most notable, however, is the clear saturation of gate error for system sizes of $\sim 100$ qubits and beyond. This data indicates that increasing system size beyond this point does not lead to lower accuracy in terms of gate errors or local fidelity of the tensor network. It should be noted that this saturation trend seems consistent with the saturation trend that appears to be emerging in the  ground-truth-based errors plotted in Fig.~\ref{fig:DWaveErrorvsSystemSize}B of the main text and thus provides strong evidence that fixed accuracy results can be obtained with linear resources for a fixed annealing time.
\par In fig.~\ref{fig:ErrorAnalysis}B we show how the deviation between the BP and first-order loop-corrected results for third nearest neighbor correlators ($d = 3$) varies with system size. Again, after a short initial increase with system size, the loop correlations appear to saturate with system size, suggesting that at fixed annealing times loop effects are stable with increasing system size. This means that we do not require increasing orders of loop corrections with system size to compute a correlator to a desired accuracy.

\par Finally, in fig.~\ref{fig:Timing} we provide walltimes for $3$D diamond simulations as a function of system size. In fig.~\ref{fig:Timing}A we provide timings for CPU-based simulations of the BP-based time evolution, demonstrating clear linear scaling with system size for fixed bond dimension and annealing time. Meanwhile, in fig.~\ref{fig:Timing}B we provide timing data for measuring a single randomly selected $z-z$ correlator on both a CPU and GPU for several bond dimensions. Again, all results show clear linear scaling with system size. Notably, there is a dramatic speedup obtained by using a GPU to take the measurement at larger bond dimensions, reaching a $10 \times$ speedup for $\chi = 32$ on the given diamond lattices. Both the contractions necessary to converge the BP message tensors and contract the necessary loops are performed on the respective hardware. 

\subsubsection*{Details of Kibble Zurek collapses}
In fig.~\ref{fig:DWaveConvergence}B of the main text we show collapse of both the spin glass order parameter and distance-dependent correlation function for diamond and cylindrical lattices respectively. Here we provide further details on the collapse performed and how we determine the Kibble-Zurek exponent $\mu$ and non-trivial fitting parameter $r$.

In the cylindrical case, a cylindrical tensor network approach is used, with a belief propagation-based time evolution protocol of the Trotterized circuit implemented with a maximum bond dimension $\chi_{\rm BP} = 32$. The final state is truncated down, with BP, to bond dimension $\chi = 8$ and MPS message passing with a MPS rank of $R = 16$ is used to calculate the relevant $z-z$ correlators.

We calculate the correlation function
\begin{equation}
    C(d) = \overline{\vert \langle \sigma^{z}_{i} \sigma^{z}_{j} \rangle - \langle \sigma^{z}_{i} \rangle \langle \sigma^{z}_{j} \rangle \vert }_{{\rm dist}(i,j) = d},
    \label{Eq:SMCorrelationFunction}
\end{equation}
where the average is taken over ten disorder realizations and ${\rm dist}(i,j)$ is defined as the length of the path (i.e.~the Manhattan distance) between the qubits $i$ and $j$. We only consider qubits that lie along the same column in the cylinder, ignore the qubits at the end of each column, and ensure one of the qubits in the correlator is in the center of the column to minimize boundary effects. We find that, for a given annealing time and cylinder size, the correlation function fits a compressed exponential  function  \cite{Trachenko2021, Wu2018} $C(d) \sim a_{0}\exp(-a_{1} d^{\alpha})$ with $\alpha > 1$ over moderate distances ($d < 10$).  We observe $\alpha \sim 1.3$.

Upon rescaling the distance as $\tilde{d} = t_{a}^{-\frac{1}{\mu}}d$ where $\mu$ is the Kibble-Zurek exponent we then observe the anticipated collapse of the correlation function \cite{Keesling2019}. The Kibble-Zurek exponent was very recently estimated in Ref.~\cite{Massimo2024} with extensive Monte Carlo simulations of the corresponding classical model in $2$ + $1$D. A value of $\mu = 2.6 \pm 0.3$ was obtained for the finite-size systems studied while scaling arguments led to an estimation of $\mu = 3.17 \pm 0.41$ in the thermodynamic limit.

Here we obtain a value for $\mu$ for a given cylinder size by finding the value where the data gives the best fit (in terms of the sum of the absolute values of the relative residuals) to the compressed exponential: $C(\tilde{d}) \sim a_{0}\exp(-a_{1} \tilde{d}^{\alpha})$ where $a_{0}$, $a_{1}$, and $\alpha$ are free fitting parameters determined from a least-squares fitting.  We see a collapse over many orders of magnitude and the curve fitting the data excellently. For the largest cylinder ($18 \times 18$) we obtain a best fit to the disorder-averaged data with $\mu \approx 2.75$ which is consistent with the finite-size prediction of Ref.~\cite{Massimo2024} and from the collapse of the Binder cumulant for the three largest cylinders in Ref.~\cite{king2024} which yielded $\mu = 2.67 \pm 0.29$.

In the diamond case, a diamond tensor network approach is used, with a belief propagation-based time evolution protocol of the Trotterized circuit implemented with a maximum bond dimension $\chi = 16$. Loop corrections with $l_{\rm max} = 6$ is used to calculate the relevant $z-z$ correlators.

For each annealing time and system size, the spin glass order parameter \newline $\langle q^{2} \rangle = \frac{2}{N(N-1)}\sum_{i < j} \langle \sigma^{z}_{i}\sigma^{z}_{j} \rangle^{2}$ is estimated from a random sample of $500$ of the correlators. Data is then averaged over $5$ independent disorder realizations. We consider the range of system sizes \newline $L_{x}, L_{y}, L_{z} = (8,8,8), (8,8,12), (8,8,12), (10,10,16)$, with $L = \sqrt[3]{\frac{L_{x}L_{y}L_{z}}{4}}$ the cube root of the number of qubits.

Collapse is then performed by rescaling $\langle q^{2} \rangle$ with $L^{r}$ and the annealing time $t_{a}$ with $L^{-\mu}$. We use a value of $\mu = 2.85$ from the literature on 3D Ising Spin glasses \cite{king2023, king2024} and determine the non-trivial parameter $r \approx 0.17$ by identifying the value of $r$ which best collapses (i.e. minimizes the residual in a least-squares fitting) the data to a straight line in the $\left( x,y \right) = \left(\log(\langle q^{2} \rangle L^{r}), \log(t_{a} L^{- \mu}) \right)$ plane.

\subsubsection*{Offsetting Truncations During Time Evolution} 
In our simulations of the dynamics of Eq.~\ref{Eq:Hamiltonian} for the cylindrical spin-glass we use a bond dimension during the evolution of $\chi_{\rm BP} = 32$ before truncating to a lower bond dimension $\chi < 32$ to enable the use of more accurate corrections of belief propagation to measure observables (whose computational scaling is higher than uncorrected belief propagation). We find this ``overshooting'' of the bond-dimension with belief propagation based simple update highly effective in comparison to keeping the bond-dimension constant and lower throughout the simulation, i.e. identical for both the evolution and measurement phases of the simulation.
\par
\begin{figure*}[t!]
    \centering
    \includegraphics[width =\columnwidth]{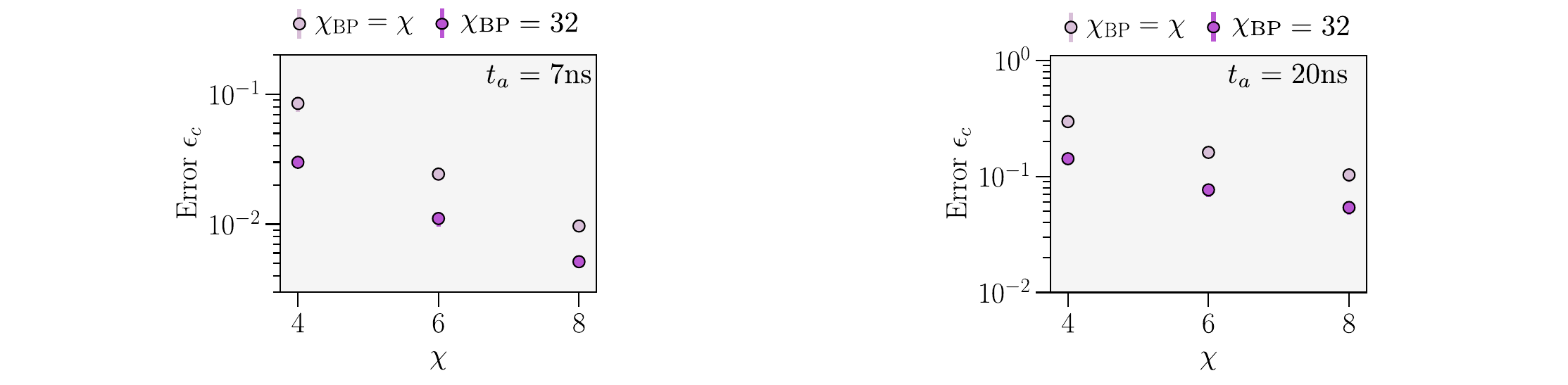}
    \caption{Error $\epsilon_{c}$---see Eq.~\ref{Eq:ErrorMetric}---from two-dimensional tensor network simulations of a quantum annealing protocol for a disordered $8 \times 8$ cylindrical spin glass. The same $N = 20$ disorder realization are used as in Ref.~\cite{king2024}. 
     Error bars correspond to double the standard error on the mean. We run a BP-based evolution protocol of the Trotterized circuit with a maximum bond dimension $\chi_{\rm BP}$ and (in the case $\chi_{\rm BP} > \chi$) truncate down, with BP, to a final state of bond dimension $\chi$ before using cylindrical boundary MPS with a MPS rank of $R = 2\chi$ to calculate $\langle \sigma^{z}_{i}\sigma^{z}_{j} \rangle \ \forall i, j$.}
    \label{fig:Overshoot}
\end{figure*}

 This is shown explicitly in fig.~\ref{fig:Overshoot} where we provide a comparison of the final error $\epsilon_{c}$ of the simulation for $\chi_{\rm BP} = 32$ and $\chi_{\rm BP} = \chi$, where $\chi$ is the bond dimension of the state at two different annealing times on the $8 \times 8$ cylindrical  lattice. The effect of maintaining a larger value of $\chi_{\rm BP}$ during the evolution is clear and leads to a large error reduction. This mode of simulation can be seen as ``offsetting'' significant truncations until the final state is obtained, where the truncation can then be done more faithfully with respect to the desired state. 

\par In the case of the three-dimensional lattices, the lowest orders of loop corrections are generally more affordable than MPS message passing with $R = \mathcal{O}(\chi)$ on the cylinder and thus we find we can obtain maintain a consistent, relatively high, bond dimension during the evolution and do not need to perform a final truncation to enable the use of loop corrections---although we still anticipate using a larger $\chi_{\rm BP}$ may lead to an improvement in the error.

\subsubsection*{Complementary Work}
Following the initial release of this paper, we became aware of a complementary effort which focussed on classically simulating the $t_{a} = 7$ns quenches on $N \leq 128$ diamond lattices with time-dependent variational Monte Carlo \cite{mauron2025challengingquantumadvantagefrontier}.
\subsubsection*{Software} All of our simulations were performed using the \textsc{TensorNetworkQuantumSimulator.jl} \cite{TensorNetworkQuantumSimulator} library, an open-source and publicly available Julia \cite{Bezanson2017} package for quantum simulation with general tensor networks built on top of \textsc{ITensors.jl} \cite{Fishman2022}.

\end{document}